# Olivine on Vesta as exogenous contaminants brought by impacts: Constraints from modeling Vesta's collisional history and from impact simulations

**Authors:** D. Turrini[1,2]*, V. Svetsov[3], G. Consolmagno[4], S. Sirono[5], S. Pirani[6]

**Affiliations:**

[1] Istituto di Astrofisica e Planetologia Spaziali INAF-IAPS, Via Fosso del Cavaliere 100, 00133 Rome, Italy.

[2] Departamento de Fisica, Universidad de Atacama, Copayapu 485, Copiapó, Chile

[3] Institute for Dynamics of Geospheres, Leninskiy Prospekt 38-1, Moscow 119334, Russia.

[4] Specola Vaticana, V-00120, Vatican City State.

[5] Graduate School of Earth and Environmental Sciences, Nagoya University, Tikusa-ku, Nagoya 464-8601, Japan.

[6] Lund Observatory, Department of Astronomy and Theoretical Physics, Lund University, Box 43, SE-221 00 Lund, Sweden.

*Correspondence to: diego.turrini@iaps.inaf.it

## Abstract

The survival of asteroid Vesta during the violent early history of the Solar System is a pivotal constraint on theories of planetary formation. Particularly important from this perspective is the amount of olivine excavated from the vestan mantle by impacts, as this constrains both the interior structure of Vesta and the number of major impacts the asteroid suffered during its life. The NASA Dawn mission revealed that olivine is present on Vesta's surface in limited quantities, concentrated in small patches at a handful of sites not associated with the two large impact basins Rheasilvia and Veneneia. The first detections were interpreted as the result of the excavation of endogenous olivine, even if the depth at which the detected olivine originated was a matter of debate. Later works raised instead the possibility that the olivine had an exogenous origin, based on the geologic and spectral features of the deposits. In this work, we quantitatively explore the proposed scenario of a exogenous origin for the detected vestan olivine to investigate whether its presence on Vesta can be explained as a natural outcome of the collisional history of the asteroid over the last one or more billion years. To perform this study we took advantage of the impact contamination model previously developed to study the origin and amount of dark and hydrated materials observed by Dawn on Vesta, a model we updated by performing dedicated hydrocode impact simulations. We show that the exogenous delivery of olivine by the same impacts that shaped the vestan surface can offer a viable explanation for the currently identified olivine-rich sites without violating the constraint posed by the lack of global olivine signatures on Vesta. Our results indicate that no mantle excavation is in principle required to explain the observations of the Dawn mission and support the idea that the vestan crust could be thicker than indicated by simple geochemical models based on the Howardite-Eucrite-Diogenite family of meteorites.

## 1. Introduction

Asteroid Vesta, explored between 2011 and 2012 by the NASA Dawn mission (Russell et al. 2012, 2013), has long been identified as the source of the Howardite-Eucrite-Diogenite (HED) suite of meteorites (McCord et al. 1970; Consolmagno and Drake, 1977; De Sanctis et al. 2012a; Prettyman et al. 2012; Reddy et al. 2012). These meteorites suggested that Vesta is a differentiated asteroid that accreted and experienced global melting (see e.g. Greenwood et al. 2014; Steenstra et al. 2016) in the first 3 Ma of the life of the Solar System (Bizzarro et al. 2005; Schiller et al. 2011; Consolmagno et al. 2015 and references therein).

Compositional models of Vesta's interior based on the information provided by the HEDs and on cosmochemical constraints (Mandler and Elkins-Tanton 2013; Toplis et al. 2013; Consolmagno et al. 2015 and references therein) indicate that eucrites and diogenites represent the main components of the upper and lower crust of the asteroid, and that the total thickness of this crust should range between 20 and 40 km (see Consolmagno et al. 2015 for a detailed discussion of the porosity of the vestan crust). A part (10-20%, Mandler and Elkins-Tanton 2013) of the lower crust was suggested to be in the form of olivine diogenite, i.e., diogenite minerals containing <40% olivine. Underneath this crust, a mantle rich in olivine (harzburgite containing 60-80% olivine, Mandler and Elkins-Tanton 2013; Toplis et al. 2013; Consolmagno et al. 2015) should extend down to an Fe-dominated core with an estimated radius of 110-140 km depending on its density (Russell et al. 2012; Mandler and Elkins-Tanton 2013; Ermakov et al. 2014; Consolmagno et al. 2015).

The Dawn mission, while confirming the global survival of the eucritic-diogenitic crust (De Sanctis et al. 2012a; Prettyman et al. 2012), revealed that the vestan surface is densely covered, plausibly to saturation level (Marchi et al. 2012; Turrini et al. 2014; Pirani&Turrini 2016), by craters of all sizes that testify to the violent collisional past of the asteroid. Moreover, the data provided by the Framing Camera (FC) onboard the Dawn spacecraft showed that the 500 km wide Rheasilvia basin, originally identified (Thomas et al. 1997) through the observations of the Hubble Space Telescope, partially overlaps an older 400 km wide basin, Veneneia (Russell et al. 2012; Marchi et al. 2012; Schenk et al. 2012).

According to numerical simulations and their comparison with Dawn's data, the impacts that created Rheasilvia and Veneneia should have reached depths of at least 40-80 km, in principle excavating the crust and exposing the harzburgitic mantle in the crater floor and walls, in the ejecta, or both (Ivanov and Melosh 2013; Jutzi et al. 2013). The data provided by the Visual and near-InfraRed spectrometer (VIR) onboard Dawn, however, ruled out the possibility of large exposures of olivine inside Rheasilvia and on the rest of the vestan surface (De Sanctis et al. 2012a; Ammannito et al. 2013). The lack of olivine signatures in the central mound of Rheasilvia, in particular, was attributed by Ruesch et al. (2014a) to the absence of significant quantities (i.e. ≥20-30%) of olivine down to the depths (≥40 km, Ivanov and Melosh 2013; Jutzi et al. 2013; Clenet et al. 2014) excavated by the impacts.

After extensive searches, two olivine-rich (50-80%) sites (Ammannito et al. 2013) associated with the post-Rheasilvia craters (Ruesch et al. 2014b) Arruntia and Bellicia were identified by VIR in the northeastern hemisphere of Vesta (see Fig. 1). The distribution of the olivine outcrops observed by VIR was described as consistent with the exposure of shallow olivine deposits (Ammannito et al. 2013).

The list of olivine-rich sites was expanded by the discovery by VIR of a second set of 11 sites (see Fig. 1) containing lower concentrations of olivine (<50% but plausibly >20-30%, Ruesch et al. 2014a), all in the eastern hemisphere except for one located in the western hemisphere inside Rheasilvia (see Fig. 1 and Ruesch et al. 2014a). Most of these sites were associated with post-Rheasilvia craters and, because of the shallow depths of the latter, were interpreted as the exposition of local plutonic deposits of olivine diogenite (Ruesch et al. 2014a). A few sites were identified on the outer part of the rim of Rheasilvia and interpreted as ballistic-deposited material originating from near-by excavated craters (Ruesch et al. 2014a).

An alternative third set of 6 less olivine-rich sites was later identified through VIR, all consisting of different outcrops with olivine concentration comprised between 25% and 50% and all but one northern of the original two (see Fig. 1 and Palomba et al. 2015). One of these sites is located inside Rheasilvia and one in the western hemisphere (Palomba et al. 2015). The two sets of 11 and 6 less olivine-rich sites do not overlap between them (Fig. 1; Ruesch et al. 2014a; Palomba et al. 2015), Arruntia and Bellicia being the only regions identified in all studies performed on the

VIR dataset. Nevertheless, in all sites identified by VIR (Ammannito et al. 2013; Ruesch et al. 2014a; Palomba et al. 2015) olivine is reported in outcrops hundreds of meters wide.

The survival of Vesta's crust and the exposure (or lack thereof) of its harzburgitic mantle has long been identified as a pivotal constraint for the study of the evolution of the asteroid belt (see O'Brien and Sykes 2011 and references therein; Consolmagno et al. 2015) and the whole Solar System (see Coradini et al. 2011 and references therein; Brož et al. 2013; Turrini, 2014; Turrini and Svetsov 2014; Consolmagno et al. 2015; Pirani&Turrini 2016). For this constraint to be valid, however, it is essential we understand the interior structure and geological history of Vesta, but the limited number of olivine-rich sites, together with the small sizes of the outcrops and the lack of high-concentration ($\geq 50\%$) outcrops inside Rheasilvia and on its central mound (Ammannito et al. 2013; Ruesch et al. 2014a; Palomba et al. 2015) cast doubts on the pre-Dawn ideas on the geophysical evolution of Vesta and the petrology of the HEDs (see Consolmagno et al. 2015 for a discussion).

The proposed association of the outcrops observed at Arruntia and Bellicia with the excavation of more surficial olivine plutons (Ammannito et al. 2013) could in principle be explained with the formation of Vesta's crust through serial magmatism in a series of shallow magma chambers instead of an extended magma ocean (Mandler& Elkins-Tanton 2013) while the lack of olivine-rich outcrops inside Rheasilvia could be the result of the deposition at depth of the olivine during the mantle crystallization, resulting in a olivine-depleted upper mantle overlying a olivine-enriched lower mantle(Toplis et al. 2013). Both these scenarios, however, have been proved not to be so straightforward in their application to Vesta once mass balance and the abundance of trace elements and rare earth elements in the HEDs are taken into account (Barrat& Yamaguchi 2014; Consolmagno et al. 2015; Steenstra et al. 2016).

The observational data from the Dawn mission recently provided a final piece of information on the subject that suggested a possible solution to this problem. A search for olivine-rich deposits performed in the dataset provided by the FC using a set of three spectral parameters (Nathues et al. 2015) resulted in the identification of a number of small outcrops, all associated with impact features. This fourth set of olivine-rich sites partly overlaps the set identified by Ruesch et al. (2014a) and partly with that reported by Palomba et al. (2015). As in the case of the previous studies (Ammannito et al. 2013; Ruesch et al. 2014a; Palomba et al. 2015), the results of Nathues et al. (2015) confirm Arruntia and Bellicia as the regions associated with stronger olivine signatures.

While the data in the visible range are less reliable than those in the infrared for the identification of olivine, the higher resolution (about a factor of three) offered by Dawn's FC with respect to that of VIR allows for more detailed information on the spatial distribution and morphology of the olivine-rich sites. Based on their results and in contrast to previous work (Ammannito et al. 2013; Ruesch et al. 2014a), Nathues et al. (2015) argued that the geologic nature and context of the vestan olivine is suggestive of an exogenous nature, a claim supported by the contemporary reanalysis of the VIR spectral data on the olivine signatures associated to Arruntia and Bellicia by Le Corre et al. (2015).

An exogenous origin of the olivine could be easily explained, from a qualitative point of view, as the natural outcome of the continuous flux of impactors on Vesta over the lifetime of the asteroid (Turrini et al. 2014) and would be supported by the results of hydrodynamic simulations of the fate of projectiles after hypervelocity impacts (Svetsov 2011; Turrini&Svetsov 2014; Svetsov and Shuvalov 2016). Recent impact experiments (Daly & Schultz 2015, 2016; McDermott et al. 2016, Avdellidou et al. 2016) have provided strong observational support to this scenario, as they confirm that, in contrast to hypervelocity (>20-30 km/s) impacts where stony projectiles undergo complete vaporization (e.g., Melosh 1989; Svetsov and Shuvalov 2016), a significant fraction of the projectile indeed survives the impact at the velocities characteristic of the asteroid belt (Daly & Schultz 2015, 2016; McDermott et al. 2016, Avdellidou et al. 2016).

The possibility of an exogenous origin for the detected olivine is indirectly supported by the comparison of its geologic context with that of the dark material on Vesta, also identified as an exogenous contaminant (McCord et al. 2012; Prettyman et al. 2012; Reddy et al. 2012; Turrini et al. 2014; Jaumann et al. 2014). As with the currently detected olivine-rich sites, the dark material appears both as surface deposits and as buried veneers, both in clear association with craters and their ejecta and as isolated spots (in some cases on topographic heights) not associated to specific craters (McCord et al. 2012; Reddy et al. 2012; Jaumann et al. 2014). As also pointed out by Ruesch et al. (2014a) for the case of the olivine deposits on Rheasilvia's rim, the existence of such isolated spots is a direct consequence of Vesta's low gravity, which allows for ballistic transport of the projectile's material on a global scale (the latter effect also enhanced by Vesta's fast rotation as pointed out by Jutzi et al. 2013). Finally, note also that the dark material is distributed non-uniformly (McCord et al. 2012; Reddy et al. 2012; Turrini et al. 2014) and shows a higher concentration of sites in a limited region north of Rheasilvia and Veneneia and a leopard-spot distribution on the rest of the vestan surface.

Previous works (Ammannito et al. 2013; Ruesch et al. 2014a) were motivated to focus on an endogenous origin scenario because of the low abundance of known olivine-dominated A-type asteroids in the asteroid belt, and the fact that impactors were expected to be completely destroyed in hypervelocity impacts (Ammannito et al. 2013, Supplementary Information; see also De Sanctis et al. 2012b for similar assumptions for the case of the vestan dark material and its delivery by carbonaceous chondritic impactors). As we reported above, however, recent work has raised doubts about the endogenous origin of the detected olivine and demonstrated that the latter assumption that the projectiles do not survive impact is not supported by either laboratory or theoretical data. Given these results, we are prompted to investigate quantitatively the role of asteroid impacts on Vesta in delivering olivine to its surface and to assess to what extent the observed features require an endogenous source or may be explained entirely in terms of exogenous contamination.

Throughout the rest of the paper, when discussing the presence of olivine on Vesta we will follow the nomenclature adopted up to now and in Fig. 1, namely: we will refer to olivine outcrops when speaking of single localized deposits of olivine on the vestan surface, while we will refer to olivine-rich sites when speaking of one or more olivine outcrops geographically or morphologically associated to a specific region or surface feature (e.g. craters) of Vesta.

## 2. Methods

Our analysis takes advantage of the contamination model developed to study the origin of the dark material observed on Vesta by Dawn as the result of the continuous secular flux of impacts of carbonaceous chondritic asteroids (McCord et al. 2012) and expanded to investigate all possible exogenous contaminants (Turrini et al. 2014). This model allows one to estimate the mass delivered to Vesta by the continuous secular flux of impactors over a selected temporal interval, taking advantage of our knowledge of the temporal evolution of the population of the asteroid belt (see Turrini et al. 2014 and Sect. 2.1). The results of the model are able to fit quantitatively several observational quantities measured by Dawn on Vesta and in the laboratory on the HEDs (see Turrini et al. 2014) including: the observed number of dark craters inside Rheasilvia (Reddy et al. 2012; Turrini et al. 2014); the amount of hydrated material measured on Vesta by the Gamma Ray and Neutron Detector (GRaND) onboard Dawn (Prettyman et al. 2012); the lack of global Fe enrichment measured by GRaND in the vestan regolith (Yamashita et al. 2013); and the Ni enrichment of howardites with respect to eucrites and diogenites found in laboratory measurements (Warren et al. 2009). Finally, the picture depicted by the results obtained by Turrini et al. (2014) with this contamination model is consistent with the results of Lorenz et al. (2007), who identified fragments of ordinary chondrites, enstatites, ureilites, and mesosiderites alongside those of carbonaceous chondrites in howardites and polymict eucrites, and gained additional observational support by the analysis of the abundances of the platinum group elements in brecciated HEDs, which recently confirmed the presence of ordinary chondrites, enstatite chondrites, and iron meteorites among the impactors on Vesta (Shirai et al. 2016).

In this work we update the contamination model used in Turrini et al. (2014) with dedicated hydrocode simulations and by extending the size range of the considered impactors. Previous studies conducted with this model focused on the role of impactors with diameter ≥1 km, while here we consider also sub-km impactors down to 200 m in diameter. This lower bound was chosen because, while the number of sub-km asteroids in the asteroid belt is subject to large uncertainties (Bottke et al. 2005; Gladman et al. 2009; Marchi et al. 2012), the crater record on Vesta is well reproduced in all different size-frequency distributions (SFDs) down to the crater diameter associated to this size of impactors (Marchi et al. 2012, 2014). In our computations we adopted the most conservative estimate (Gladman et al. 2009) for the number of sub-km asteroids among the available ones, which means that our results may underestimate the real values by about a factor of two (Marchi et al. 2012; see also Sect. 2.1). Based on the results of Le Corre et al. (2015)'s spectral modeling of the olivine signatures of Arruntia and Bellicia, we focused on two different classes of olivine-carrying impactors: A-type and S-type asteroids (see Fig. 2 and Sects. 2.1-2.3).

## 2.1. The contamination model

As mentioned above, to assess the number of different kinds of impactors on Vesta over time and the global amount of material they leave on its surface we used an updated version of the contamination model from Turrini et al. (2014). The contamination model uses the average intrinsic impact probability of Vesta with the bodies populating the asteroid belt together with a time-evolving description of the SFD of the asteroid belt to assess statistically the number and sizes of the impactors on Vesta over a given temporal interval (Turrini et al. 2014). Because of the statistical nature of the model, the intrinsic uncertainty on the estimated number of impactors $n_i$ is $\sqrt{n_i}$ as per Poisson statistics (see Turrini et al. 2014 for the discussion of the different sources of uncertainty). From the abundances of the different spectral classes of asteroids in the JPL Small-Body Database Search Engine and from the composition of the different classes of meteorites used as their analogues (McSween et al. 1991; Sanchez et al. 2014) it is then possible to estimate the frequency of specific types of impactors and the delivered abundances of specific elements or minerals.

Following O'Brien and Sykes (2011) and references therein, the number of impacts on Vesta due to impactors having diameter $D_i$ over a timespan $\Delta T$ is computed as $n_i = P_V \cdot A_i \cdot N(D_i) \cdot \Delta T$ where $N(D_i)$ is the number of asteroids having diameter $D_i$ in the asteroid belt, $P_V$ is the average intrinsic impact probability of Vesta, $A_i = (R_V + 0.5 D_i)^2$ is the cross-sectional area of Vesta and these impactors (the factor $\pi$ being included in $P_V$, see O'Brien & Sykes 2011 and references therein), and $R_V = 262.7$ km is the mean radius of Vesta (Russell et al. 2012). Strictly speaking, this equation is valid only for a population $N(D_i)$ that is constant over time: while the error in the estimated number of impacts introduced by assuming the present population of the asteroid belt as constant over the last 1 Ga is limited (of the order of 1% for asteroids with $D > 1$ km, see Turrini et al. 2014), the error introduced by such an assumption when integrating over longer temporal intervals (moving backward in time) is significantly larger (Turrini et al. 2014). It is therefore necessary to correct the population of the asteroid belt for the depletion caused by its secular dynamical evolution.

The population of large ($D > 10$ km) asteroids was estimated to have declined as $f(t) = C (t/1 \text{ year})^{-D}$ where $C = 1.0556$ and $D = 0.0834$ are constants (Minton and Malhotra 2010). Such decline is due to the chaotic diffusion of these large asteroids into the network of orbital resonances crossing the asteroid belt and their subsequent ejection from the latter (Minton and Malhotra 2010). Over the last 4 Ga (the temporal interval of validity of the previous expression for $f(t)$, Minton and Malhotra 2010) such a decline causes an integrated depletion factor $F_{4Ga} = 2$, of which about 98% occurred between 4 and 1 Ga ago. No similar estimate is available for the depletion of smaller bodies, whose rate of diffusion into the resonances would be enhanced by the Yarkovsky drift but whose population would be contemporary replenished by the fragments generated by the impacts between larger bodies in the asteroid belt. As a consequence, we followed O'Brien & Sykes (2011) and references therein and conservatively assumed the latter two effects to compensate each other and the depletion rate of the smaller bodies to be the same as that of the larger bodies. For a more detailed discussion on the temporal evolution of the population and the size-frequency distribution of the asteroid belt we refer interested readers to the review by O'Brien & Sykes (2011).

Using the depletion factor $F_{4Ga}$, it is possible to define the primordial population of asteroids 4 Ga ago as $N_{prim}(D)=F_{4Ga} \cdot N_{now}(D)$, $N_{now}(D)$ being the present population of asteroids, and express the instantaneous asteroid population at time $t$ during the last 4 Ga as $N(D)=f(t) \cdot N_{prim}(D)$. For the asteroids with diameter $D$ greater than 1 km in $N_{now}(D)$ we followed Turrini et al. (2014) and adopted the present SFD $N(D>1\ km)$ described by Bottke et al. (2005). For sub-km asteroids we adopted the conservative estimate of their abundance provided by Gladman et al. (2009) down to 200 m in diameter using the magnitude-to-diameter conversion of Marchi et al. (2012). The difference (in excess) between the number of sub-km impacts on Vesta so computed and the (larger) one obtained using the sub-km SFD provided by the theoretical results of Bottke et al. (2005) is of about a factor of two.

The value of the average intrinsic impact probability $P_V$ of Vesta with any of the asteroids in the main belt, $P_V=2.72 \times 10^{-18}\ km^{-2} \cdot yr^{-1}$, was computed by O'Brien and Sykes (2011) averaging over all possible encounters (or lack thereof) of Vesta with the asteroids larger than 30 km in diameter (the size cut-off being imposed by observational completeness), assumed as dynamically representative of the whole asteroid population (O'Brien and Sykes 2011 and references therein). The comparison between the values of $P_V$ computed using different techniques (O'Brien and Sykes 2011, Marchi et al. 2014) shows the uncertainty on $P_V$ is 5-10%, less than that caused by other sources of uncertainty (see above and Turrini et al. 2014). As shown in Fig. 2 the distribution of the A-type objects identified so far can be considered, as a first approximation, as homogenous throughout the asteroid belt. More properly, the distribution of A-type objects in the asteroid belt appears homogeneous enough that their impact probability with Vesta should not differ significantly from the adopted value of $P_V$ (note that, as shown by O'Brien & Sykes 2011, it takes the much more tight clustering in orbital elements around Vesta of the Vestoids to significantly increase their impact probability with Vesta with respect to the average value of $P_V$ we are adopting). The S-type asteroids, on the contrary, show a more marked concentration into the inner and middle asteroid belt (see Fig. 2). Using the value of $P_V$ from O'Brien and Sykes (2011) can result in underestimating their impact frequency on Vesta: as we are assessing whether the impacts can bring enough exogenous contaminants to explain the observed features under conservative assumptions, however, underestimating the flux of S-type impactors does not undermine the results.

For a given population of asteroids having diameter $D_i$ the number of impacts $n_i$ on Vesta over a temporal interval $\Delta T$ can be computed as $n_i=\int P_V \cdot A_i \cdot N_{prim}(D_i) \cdot f(t)dt$, where the integral is evaluated between $t_1=0$ and $t_2=\Delta T$ as a sum over discrete timesteps $\Delta t=1000$ years (Turrini et al. 2014), i.e. $n_i=\sum_j P_V \cdot A_i \cdot N_{prim}(D_i) \cdot f(t_0+j\Delta t) \cdot \Delta t$ where $t_0=1$ Ma (Minton and Malhotra 2010) and $0 \leq j \leq \Delta T/\Delta t$. The number of A-type impactors is computed as $n_i=\sum_j F_A \cdot P_V \cdot A_i \cdot N_{prim}(D_i) \cdot f(t_0+j\Delta t) \cdot \Delta t$ where $F_A$ is the fractional abundance of A-type impactors in the asteroid belt (see Sect. 2.2). A similar equation holds for S-type impactors substituting $F_A$ with $F_S$, where $F_S$ is the fractional abundance of S-type impactors (see Sect. 2.2). If $\rho=2400$ kg·m$^{-3}$ is the average density of the impactors (Turrini et al. 2014) and $R_m$ is the retention efficiency of Vesta (i.e. the average mass fraction of the impactor retained by the asteroid, see Sect. 2.3) in the impact simulations, the mass of olivine accreted by Vesta is $m_{i,oliv}=((\pi/6) \cdot D_i^3 \cdot \rho) \cdot R_m \cdot n_i \cdot R_{oliv}$ where $R_{oliv}$ is the fraction of the projectile's mass represented by olivine (see Sect. 2.2).

**2.2. Olivine-carrying asteroids**

A-type asteroids possess spectra dominated by olivine (>70%, Sanchez et al. 2014), which suggest that they are fragments of the mantle of differentiated asteroids that underwent catastrophic disruption (Burbine et al. 1996). The JPL Small-Body Database Search Engine reports only $n_A=12$ of A-type asteroids in the asteroid belt out of a total number of asteroids with a defined spectral type $n_{ST}=1718$ (Fig. 2). Size estimates are currently available only for the 6 brightest (absolute magnitude H<12) A-type asteroids, all of which have diameters $10<D \leq 60$ km. The other, fainter six (H>12), plausibly have diameters $D \leq 10$ km, as supported by the size estimates of two A-type asteroids located outside the asteroid belt that have $D \approx 5$-6 km and H=14. Recent surveys have announced the discovery of ~20 additional A-type asteroids (DeMeo et al. 2013), suggesting that the present sample is not observationally complete (especially at smaller diameters).

In our model we estimated the plausible total population of A-type asteroids taking advantage of the current understanding of the asteroid belt. The SFD of the asteroid belt has been in collisional equilibrium over the last 4 Ga and its population at D<100 km is the product of the disruption of larger asteroids (Bottke et al. 2005). Due to the collisional origin of the A-type asteroids and their estimated sizes, it is not unreasonable to assume that their SFD is characterized by the same differential slope as the population of all asteroids with D<100 km. Assuming that the spectrally-characterized sample of asteroids $n_{ST}$ is representative of the whole population of the asteroid belt (Turrini et al. 2014), the fraction of A-type impactors among all possible impactors on Vesta should be $F_A=n_A/n_{ST}=7.0\times10^{-3}$ or 0.7%. We did not include in this estimate the announced new discoveries as no further data on them are currently available in the literature: this means we again adopted the most conservative case and the number of A-type impactors on Vesta, in the limits of our assumptions, may be underestimated by about a factor of three. The adopted fractional olivine content of A-type asteroid in the contamination model is $R_{oliv,A}=1$ ( or 100 wt%).

As indicated by the results of Le Corre et al. (2015), A-type asteroids are not the only potential carriers of olivine on Vesta. Olivine constitutes a significant fraction of ordinary chondrites (McSween et al. 1991, Dunn et al. 2010), which we adopted as our template for the composition of S-type asteroids (Fig. 2 and Turrini et al. 2014). From the study of the normative mineralogies of a sample of 94 ordinary chondrites, McSween et al. 1991 reported that, on average, olivine represents 35.0±3.9 wt% of H chondrites, 44.8±3.0 wt% of L chondrites, and 51.9±5.6 wt% of LL chondrites. Using X-ray diffraction on a sample of 48 equilibrated ordinary chondrites, Dunn et al. (2010) measured the average olivine contents of the different petrographic sub-groups of H, L and LL ordinary chondrites: from their results one can derive a weighted average abundance of olivine of 33.1±2.4 wt% for H chondrites, 42.1±0.9 wt% for L chondrites and 51.1±1.1 wt% for LL ordinary chondrites. The average olivine content of the S-type impactors on Vesta can be estimated from the previous values using the abundances of H, L and LL chondrites among the S-type asteroids in the asteroid belt. These abundances, however, are currently a matter of debate. Looking at the abundances of the different classes of ordinary chondrites in the Meteoritical Bulettin Database, H chondrites represent ~41% of all the falls, L chondrites ~47% and LL chondrites ~11%. Among the Near-Earth Asteroids (NEAs), however, H chondrites represent only ~15% of NEAs, L chondrites ~10% and LL chondrites ~60% (Vernazza et al. 2008; de Leon et al. 2010; Dunn et al. 2013). Using the meteoritic abundances in computing the weighted mean, the average olivine content of S-type impactors would range between ~39 and ~42 wt% using the values from McSween et al. (1991) and Dunn et al. (2010), respectively. Using instead the abundances derived from the NEAs, the average olivine content of S-type impactors would range between ~47 and ~48 wt% using the values from McSween et al. (1991) and Dunn et al. (2010) respectively. As our nominal fractional olivine content for S-type impactors we therefore adopted the intermediate value of 45 wt% ($R_{oliv,S}=0.45$). Using the number of known S-type asteroids in the asteroid belt from the JPL Small-Body Database Search Engine ($n_S$=563) and assuming the sample $n_{ST}$ as representative of the whole asteroid population, S-type asteroids represent $F_S=n_S/n_{ST}=0.33$ or 33% of all impactors on Vesta.

The olivine content in carbonaceous chondrites, the template for the composition of C/D-type asteroids proposed as the source of the dark material and representing about 22% of the impactors on Vesta (see Turrini et al. 2014 and references therein) can in some cases be larger than that of ordinary chondrites (see e.g. Le Corre et al. 2015 and references therein). However, in our analysis we did not consider the contribution of carbonaceous chondrites to the delivery of olivine to Vesta for the following reasons. First, with the exception of three sites identified by Palomba et al. (2015), the albedo features of the olivine-rich sites identified so far are not consistent with those of the dark material (i.e. the olivine deposits are much brighter than the dark material). For three sites associated to the craters Severina, Manilia and Caparronia, Palomba et al. (2015) reported the possible presence of olivine associated to albedo values lower than the cut-off albedo (0.43) they used to identify the dark material. As discussed by Palomba et al. (2015), these detections are the least certain among their set of olivine-rich sites since the low albedo of the outcrops identified as olivine-rich can alter the behavior of the adopted spectral indicators and result in false positives. More generally, as pointed out by Palomba et al. (2015) the low albedo of the dark material can mimic and mask the presence of olivine in the dark material itself, making any identification unreliable. Finally, olivine can be removed and transformed into other minerals due to aqueous alteration and the bulk of the dark material on Vesta appears to have originated from aqueously altered bodies. Specifically, on one hand the bulk of the carbonaceous chondritic clasts in the HED meteorites is associated to CM and CR chondrites (Zolensky et al. 1996; Lorenz et al. 2007) and, on the other, the largest part of the dark material on Vesta (more than 90% according to Palomba et al. 2014) appears to contain hydroxylate materials (De Sanctis et al. 2012b; Prettyman et al. 2012; Palomba et al. 2014; see also Nathues et al. 2014 for the detection by Dawn's FC of serpentine as a product of aqueous alteration in association to the dark material). For further details on the quantitative assessment of the flux of the dark material we refer the interested readers to Turrini et al. (2014).

**2.3. Impact simulations and retention efficiencies**

To assess the amount of olivine delivered by impacts, we performed a set of 3D numerical simulations of impacts of dunite and granite projectiles onto Vesta. We used a modified version (Svetsov 2011, Turrini and Svetsov 2014) of the numerical hydrodynamic method SOVA (Shuvalov 1999; SOVA is an acronym for Solid-Vapour-Air, as the code is designed for simulations of multi-material, multi-phase flows) that adapts the hydrodynamic equations and method to a 3D spherical system of coordinates, with a central gravitational field constant in time, and includes the effects of dry friction (Dienes and Walsh 1970). The equations with dry friction are similar to the Navier-Stokes equations and do not violate hydrodynamic similarity, but depend on a dimensionless coefficient of friction for which we adopted a value of 0.7 typical for rocks and sand.

In this work we used SOVA to solve numerically the equations of motion on a numerical grid of 250×100×225 cells over azimuth, polar angle, and radial distance respectively; we assumed bilateral symmetry to model only the half-space in the zenith direction (for this reason the distribution of ejecta shown in the Figs. 5-6 is symmetrical with respect to the plane). The impact velocity vector lies in the reference plane that passes through the origin of the coordinates and is orthogonal to the zenith. The size of the cells is 1/40 of the diameter of the projectile around the impact point and increases to the antipodal point and to the radial boundaries located at distances of about 10 vestan radii. The target, Vesta, had a mass of $2.59 \times 10^{23}$ g (Russell et al. 2012), radius of 260 km and was in equilibrium and at low temperature: the radius of its core was assumed to be 110 km (Russell et al. 2012) while the thickness of the basaltic crust was set to 23 km (i.e. that of the eucritic layer, Consolmagno et al. 2015).

To determine the behavior and properties of the target and the projectile we used the ANEOS equations of state (Thompson and Lauson 1972). The ANEOS equations are widely used in impact simulations but they offer only a restricted number of materials with reliable equations of state. Among these are dunite, granite, quartz and other rocks, but there are no equations of state for real targets and impactors of planetary science interest, which need to be approximated with available materials. We assumed the mantle of Vesta as consisting of dunite and the crust of the asteroid of granite: we used the ANEOS equations of state for mantle and crust with input data (i.e., about 35 variables describing properties of a given material like mass fractions of elements, melting temperatures, characteristic densities of phase transitions, etc.) from Pierazzo et al. (1997) and Tillotson's equation of state for the iron core (Tillotson 1962). For the impactors, we adopted dunite as our template for A-type asteroids and granite for S-type asteroids. Dunite has an initial density at zero pressure of 3.32 g/cm$^3$ while the density of granite is 2.63 g/cm$^3$. We assumed that the target is nonporous, so we calculated the lower limit of projectile mass embedded in the target after the impacts because target porosity increases the survivability of a projectile due to the lower shock pressures experienced by the projectile (see e.g. Avdellidou et al. 2016).We chose as our reference impact scenario (Fig. 3) that of an impactor of 1 km in diameter hitting Vesta at the average impact speed of 4.75 km/s (O'Brien and Sykes 2011) and at the average impact angle of 45° (Melosh, 1989). Our previous studies (McCord et al. 2012; Turrini et al. 2014) with the contamination model showed that the adoption of these average values allows for a reasonable reproduction of the global properties of exogenous contaminants on Vesta. For these values about 53% of the mass of A-type projectiles ($R_m$=0.53) and 47% of the mass of S-type impactors ($R_m$=0.47) remain on the vestan surface (Figs. 4-5), mostly inside the craters but also partly distributed on a global scale (Figs. 4-6).

To test the dependence of our results on the chosen projectile properties, we ran additional simulations with different projectile diameters and compositions. We simulated impacts occurring at velocities ranging between 1 and 10 km/s (O'Brien and Sykes 2011; see Fig. 4b for the impact velocities of 2 and 8 km/s), the impact of a projectile of 10 km in diameter (Fig. 4a) and the impact of a differentiated projectile of 1 km in diameter with an iron core 500 m in diameter (Fig. 4a). The purpose of the latter choice was mainly to estimate the numerical variations of the results depending on the composition of the projectile. Specifically, while the composition and size of the impactor could make it an analogue for a stony-iron meteoritic impactor (e.g. a fragment generating from the core-mantle interface of a destroyed differentiated asteroid), the adopted distribution of the materials is not necessarily realistic (more realistic ones would have the two materials intermixed as in pallasites or occupying two different hemispheres of the projectile). As shown in Figs. 4-5, the differences in the results due to the composition and impact velocity of the projectiles between these test cases and our reference case are on the order of 10-20%, far smaller than those due to the other sources of uncertainty (the numbers of sub-km and A-type impactors and the intrinsic uncertainty of the contamination model, see Sect. 2.2 and Turrini et al. 2014). The size of the projectile plays a somewhat more significant role, as it affects the distribution of the projectile material on the vestan surface. Our simulations show that for impactors with D≈1 km (Fig. 4) the contaminants concentrate inside the craters (~30% of the original projectile mass) and on their immediate surroundings (≤ 4 crater radii, ~5-10% of the original projectile mass), with a fraction (~10-20% of the original projectile mass) distributing globally on the hemisphere centered on the impact crater; the remaining projectile mass is lost as it escapes Vesta's gravity. This means that about 70% of the retained projectile mass remains inside or near the crater and about 30% distributes globally on Vesta. For larger impactors (D≈10 km) ~80% of the retained projectile mass (i.e. the contaminants) remains inside or near the crater, the rest distributes globally on the vestan surface (Figs. 4-6). A small fraction (<10% of the retained projectile mass) of the contaminants can always reach the hemisphere that is antipodal with respect to the crater (Figs. 4-6). As mentioned above, in our simulations we restricted the range of velocities to a maximum of 10 km/s because collisions at higher speeds are quite rare among the asteroids in the main belt (O'Brien & Sykes 2011). However, some portion of the projectile is retained also at higher velocities (on average at 20 km/s ~7% of the projectile is retained) and only after collisions with velocities higher than 25 km/s almost all the mass of a dunite projectile escapes the target (Svetsov 2011). For impacts occurring at 45°, the fragments of dunite projectiles remain in solid state up to collision velocities of 10 km/s, although at 10 km/s they can be heated to temperatures above 1000 K (Svetsov and Shuvalov 2015). About half of the mass of dunite projectiles will melt at impact velocities of 14 km/s. As the vast majority of the impacts on Vesta occurs at velocities lower these values, we do not expect the impact velocity to significantly affect the spectral appearance of the olivine-rich outcrops (e.g. by altering the olivine due to impact melting).

## 3. Results

The instruments onboard Dawn capable of detecting olivine on Vesta, FC and VIR, are sensitive to the composition of the topmost <1 cm layer of its surface. If the detected olivine is exogenous and is delivered by impacts, the highest concentrations should be due to impacts recent enough not to be affected by significant ejecta blanketing (Turrini et al. 2014), impact gardening (Pieters et al. 2012) or mass wasting (Jaumann et al. 2012). According to impact simulations

(Ivanov and Melosh 2013; Jutzi et al. 2013), the formation of Rheasilvia caused the last global ejecta blanketing of Vesta about 1 Ga ago (Marchi et al. 2012): no major impact/blanketing event occurred since then (Marchi et al. 2012; Turrini et al. 2014).

As said above, the two most olivine-rich sites identified on Vesta are indeed associated with two post-Rheasilvia craters: Bellicia, whose estimated absolute age ranges between 630 Ma and 1083 Ma (in any event more recent than the Rheasilvia-forming event, Ruesch et al. 2014b), and the much younger Arruntia, whose absolute age ranges between 2 and 15 Ma (Ruesch et al. 2014b). Based on their respective diameters of 41.7 km and 10.5 km (Gazetteer of Planetary Nomenclature, http://planetarynames.wr.usgs.gov/nomenclature/SearchResults?target=VESTA&featureType=Crater,%20craters), these craters were plausibly excavated by impactors of ~4 and ~1 km in diameter, respectively. Our model predicts that over the last 1 Ga Vesta should have undergone 2.0±1.4 impacts of A-type asteroids ≥1 km in diameter (the ones needed to form Bellicia and Arruntia) and ~14±4 impacts of sub-km A-type asteroids (capable of producing craters between ~1 and ~10 km in diameter). Given that the surface of the Rheasilvia impact basin (Schmedemann et al. 2014) accounts for ~12.5% of the whole vestan surface, the fraction of the total number of A-type impactors expected to fall inside is (coincidentally the same) ~2.0±1.4.

Impacts of A-type asteroids alone, from Rheasilvia's formation to the present, are therefore able to supply, within 1σ, enough events to explain the two most olivine-rich sites in the north-eastern hemisphere of Vesta, all the additional less olivine-rich sites, and the possible presence of 1-3 of them inside Rheasilvia. As our results can underestimate the real number of events by a factor of a few due to our conservative assumptions (using the highest values for the abundances of A-type and sub-km asteroids would result in ~6 times as many events), post-Rheasilvia impacts of A-type asteroids not only can explain all currently identified features but leave some margin for inefficient delivery (i.e. not all impacts leave detectable outcrops of olivine, e.g. by depositing the olivine over the dark material), removal (i.e. outcrops are buried by ejecta and landslides, hit by later impactors or diluted by impact gardening), and for new discoveries of olivine-rich sites.

The specific characteristics of the vestan collisional environment, however, offer another possibility to link the detected olivine patches on the asteroid to the impact of A-type asteroids. Because of its location in the asteroid belt, the crater production rate of Vesta (in units of craters larger than or equal to 1 km in diameter per $km^2$) is about 20 times higher than that of the Moon (Schmedemann et al. 2014; O'Brien et al. 2014), while at the same time the smaller radius of Vesta makes its surface about 2% that of the Moon. As a consequence of the combination of these two factors, ejecta blanketing should be about three orders of magnitude more effective on Vesta than on the Moon: according to the results of Turrini et al. (2014), the cumulative ejecta blanketing caused by impacts on Vesta over the last 2.6 Ga would be enough to affect the surface of the asteroid to a global level even without including the effects of Rheasilvia's and Veneneia's formation.

When one includes in the picture the global blanketing events caused by the formation of Veneneia and Rheasilvia, it does not appear implausible that the currently detected olivine-rich sites could also be the outcome of pre-Rheasilvia impacts of A-type asteroids that were buried by the ejecta of subsequent impacts (thus preserved from dilution by gardening) and re-exposed or excavated by more recent cratering events. In this scenario, which would fit the geo-morphological interpretation of the olivine sites made by Ammannito et al. (2013), the currently detected olivine-rich sites would represent a fraction of the olivine originally delivered to Vesta (i.e. that not destroyed by later impacts and buried deep enough not to have been diluted beyond our detection capability).

Theoretical results validated against the observations of the Dawn spacecraft (Turrini et al. 2014; Pirani & Turrini 2016) suggest that the collisional evolution of Vesta over the last 4 Ga caused the saturation of the surface of the asteroid with craters and stripped away most of the previously deposited contaminants: the exogenous contaminants presently mixed in the vestan regolith (among which the olivine) should then have been delivered over this timespan. Our contamination model indicates that, if A-type asteroids represented a more or less constant fractional population of the asteroid belt over the last 4 Ga, up to 51±9 impacts of A-type asteroids larger than 200 m could have fallen on Vesta between the Late Heavy Bombardment and the formation of Rheasilvia. These values indicate that, in this scenario, about one impact out of three (one out of four when considering the contribution of the last 1 Ga) should be able to create detectable outcrops in order to fit the number of detected sites.

Alongside A-type asteroids, S-type impactors would also deliver olivine to Vesta, albeit mixed with other minerals: over the last 1 Ga, ~700±30 S-type impactors should have fallen on the asteroid. If S-type asteroids are indeed responsible for the detected olivine-rich sites as suggested by Le Corre et al. (2015), these values indicate that only ~2% of said impacts needs to create olivine-rich outcrops with olivine concentrations within Dawn's detection capabilities in order to match the observational data. Notwithstanding its inefficiency in creating local olivine enrichments, the large flux of S-type impactors could in principle leave global signatures in the composition of the vestan regolith. We tested whether our results are consistent with Dawn's data also from this point of view.

As we mentioned above, impacts on Vesta over the last 4 Ga saturated its surface with craters and stripped away most of the older contaminants, while at the same time delivering the bulk of the exogenous contaminants presently mixed in its regolith. Over this timespan Vesta would have been hit by ~3200±60 S-type impactors with diameters greater than 200 m, associated with a mass flux of $5.2 \times 10^{15}$ kg and a total delivery of olivine of $2.4 \times 10^{15}$ kg. Because of their small number, the contribution of the A-type impactors to this mass flux is negligible: the 67 A-type impactors expected to fall on Vesta over the last 4 Ga that we mentioned above would cumulatively deliver only $1.5 \times 10^{13}$ kg of olivine.

The spectral/compositional global signature this olivine would create depends on its degree of mixing in the regolith. The average regolith thickness over Vesta is unconstrained but local values associated to degraded craters suggest it can locally reach several hundred meters of depth (Jaumann et al. 2012). If we assume a minimum average regolith depth of 100 m with a density of 2000 kg/m$^3$ (i.e. a regolith layer with mass of $1.7 \times 10^{17}$ kg, McCord et al. 2012, Turrini et al. 2014), once homogeneously mixed the exogenous olivine would produce an average abundance of ~1.2 wt%, far too small to be detectable. As the mass flux of olivine associated to A-type impactors is two orders of magnitude smaller than that associated to S-type impactors, their contribution to the global olivine enrichment of the vestan regolith is marginal and can be neglected.

Observational data, however, indicate that the vestan regolith is not homogenously mixed in the vertical direction. The dark features observed on Vesta show a larger presence of contaminants near the surface than at depth and a clustered distribution of the buried contaminants as localized deposits (McCord et al.2012; Reddy et al. 2012; Jaumann et al. 2014). In addition, the comparison between the GRaND's hydrogen measurements (Prettyman et al. 2012) and the estimates of the contamination model (Turrini et al. 2014) indicates that even if only ~10% of the exogenous hydrogen-rich contaminants remains in the topmost 1 m of the regolith, this would be enough to match GRaND's observations, independent of the fate of the remaining 90% (i.e. removed by later impacts or simply buried deeper). If we assume the exogenous olivine's vertical distribution in the regolith mimics the one estimated for hydrogen, i.e. ~10 wt% of the olivine from S-type impactors remains in the topmost 1 m, the impacts would produce a ~12 wt% abundance in this 1 m-thick layer. While an order of magnitude larger than our first estimate, this value is still below the current detection threshold (Ruesch et al. 2014a; Palomba et al. 2015).

The experimental results of Daly and Schultz (2015) give an average retention efficiency of the projectile material of about 17% in the regions near the crater and in its immediate surroundings. According to our simulations, these regions would account for about 70% of the total delivered projectile material (see Sect. 2.3 and Fig. 4): scaling to the results of our impact simulations, the total retention efficiency should round up to about 24%, i.e. about a factor of two lower than in our simulations. Adopting the retention efficiency scaled from impact experiments, therefore, the olivine enrichment of the topmost 1 m of the vestan regolith would not exceed 6%. However, it should be pointed out that the results of these laboratory experiments present a lower limit for the retention efficiency because the target gravity was not simulated. As follows from scaling laws (Melosh 1989), the Earth's gravitational field is insufficient for the simulation, in small-scale impact experiments, of the gravitational effects in the larger impacts we are considering here for the case of Vesta.

These global estimates over 4 Ga are not valid for Rheasilvia, as the basin's formation removed all previously delivered contaminants (also those buried at depth), but dedicated calculations show that the flux of S-type impactors after Rheasilvia's formation would not be capable of leaving a detectable global olivine signature in the basin. Post-Rheasilvia S-type impactors would deliver $4.1 \times 10^{13}$ kg of olivine inside the basin: even in the extreme case all this olivine remained in the topmost 1 m of regolith, the resulting global enrichment would be of ~15% (7% should we adopt the retention efficiency scaled from the impact experiments of Daly & Schultz 2016), again below the detection threshold of current surveys (Ruesch et al. 2014a; Palomba et al. 2015).

**4. Discussion and conclusions**

These results show that the collisional evolution of Vesta would naturally provide sufficient exogenous olivine to explain all identified features as the result of impact contamination without violating Dawn's constraint on the lack of global olivine signatures. The overall efficiency of the collisional delivery process would depend on the timeframe of the delivery (i.e. before or after the formation of Rheasilvia) and on the carrier of the olivine: A-type asteroids, S-type asteroids or both.

If A-type asteroids alone are responsible for the detected outcrops, the delivery process should be characterized by a reasonably high efficiency: depending on the temporal interval over which the olivine is delivered, no less than 25% of the impacts should leave behind one or more detectable outcrops (unless sub-km A-type asteroids are more abundant than in our reference case, as discussed above). This is not implausible given the high olivine content of A-type asteroids and the small sizes of the outcrops, but requires that the removal/dilution of outcrops after their formation/excavation by impacts should be limited. This requirement, however, is consistent with the geologically young age of the olivine-rich sites.

The case of S-type impactors is the opposite, as no more than 2% of the impacts should create detectable outcrops in order to reproduce Dawn's observations. This also is not implausible, considering the lower olivine concentration of S-type asteroids: most olivine outcrops created by their impacts would be characterized by olivine concentrations lower than 50% from the start and only a limited dilution in the regolith by gardening would be required to bring their olivine concentration below detectability. In this end case, however, outcrops with olivine concentrations >50% would be a rare stochastic outcome.

The most likely and natural scenario, however, is the one where both types of asteroids contribute to the olivine delivery and are responsible for the creation of olivine-rich sites, with the A-type asteroids plausibly causing the creation of those sites with olivine concentration about 50% and S-type asteroids contributing to the creation of the low concentration ones. The contribution from both kinds of impactors would relax the issues associated to the removal/burial of the olivine and the creation of outcrops with concentrations >50%. In this scenario our results would leave ample room for inefficiencies in the formation process of the olivine outcrops and/or for the discovery of more olivine-rich sites in the future.

**4.1. Implications for the study of Vesta and of the asteroid belt**

Once they are considered along with the general lack of a marked (>50%) olivine signature on most of Vesta's surface, particularly inside Rheasilvia and on its central mound, our results do not leave room for the presence of any significant abundance of olivine from any other source, such as deep excavation of Vesta's interior. Instead, our results suggest that significant quantities of olivine from the mantle of Vesta were not excavated nor exposed in any point of the vestan surface, which in turn would suggest that the crust of Vesta must be thicker than indicated by previous geochemical models. These models had assumed that Vesta was formed from material with chondritic proportions of the major rock-forming elements: our results therefore would support the idea that either Vesta formed with a non-chondritic composition (Clenet et al. 2014; Consolmagno et al. 2015), or some major event altered the composition of the HED parent body to Vesta's current composition during the early life of the asteroid (Consolmagno et al. 2015).

It is interesting to note that, should future data support instead the case for an endogenous origin of the detected olivine, our results would pose an upper limit to the abundance of A-type impactors onto Vesta and, consequently, of A-type asteroids in the asteroid belt, implying that they account for less than 0.7% of its present population. However, data on iron meteorites suggest that differentiated asteroids were more abundant than indicated by the present population of A-type asteroids (Burbine et al. 1996; see Consolmagno et al. 2015 for a recent detailed discussion) and the JPL Small-Body Database Search Engine reports $n_M$=37 M-type asteroids (thought to be the remnants of the cores of differentiated planetesimals) in the asteroid belt out of the total number of asteroids with a defined spectral type $n_{ST}$=1718 (i.e. a factor of 3 higher than the number of A-type asteroids). Moreover, 3D geophysical simulations suggest that most asteroids bigger than ~20 km that formed in the first 1.5 Ma from the condensation of CAIs should have been able to differentiate and form olivine mantles (see e.g. Golabek et al. 2014).

It has been suggested (Burbine et al. 1996) that collisional processes could be more efficient in shattering the surface and mantle material than the core material of differentiated planetesimals, which would be consistent with the fact that all M-type asteroids have D>30 km whereas only 3 A-type asteroids out of 12 fulfill the same condition. Once grinded towards the lower end of the size-frequency distribution of the asteroid belt by collisions, such remnants of the surface and mantle of differentiated planetesimals would be more efficiently removed from the asteroid belt itself by the combined effect of Yarkovsky drift and chaotic diffusion into the resonances than their core counterparts. A failure to see exogenous olivine on Vesta would then imply that such preferential removal of the mantle material has been more efficient than the observational data presently suggest, i.e. that the A-type asteroids currently known do not represent the high-end tail of a significantly larger population or, in other words, that the present population of km-sized A-type asteroids do not follow the size-frequency distribution that would be expected as resulting from collisional processes alone. However, as discussed in O'Brien & Sykes (2011) and references therein, the collisional environment of the asteroid belt over the last 4 Ga would not be able to produce such an efficient grinding and consequent removal: the source of the depletion would therefore date back to the first few hundred Ma of the life of the Solar System.

If the bulk of these now-extinct differentiated asteroids had diameters smaller than ~200 km (Burbine et al. 1996; Turrini et al. 2012), the mass loss they would have suffered due to impacts during the primordial bombardment triggered by Jupiter's formation (in the cases of moderate, ≤0.5 au migration or no migration of the giant planet, Turrini et al. 2012) and the subsequent phase of dynamical clearing of the inner Solar System(see O'Brien & Sykes 2011,Coradini et al. 2011 and references therein) could be sufficient to cause the efficient early shattering and removal of the bulk of these olivine-dominated asteroids (see Turrini 2014 and Turrini and Svetsov 2014 for a discussion of the survival of Vesta's basaltic crust and its match with HEDs in this scenario). If, instead, most of the differentiated bodies were in the form of larger planetesimals (200-600 km in diameter, Burbine et al. 1996; Turrini et al. 2012) that undergo less easily catastrophic disruption (see O'Brien & Sykes 2011 and references therein), the efficient removal of these differentiated asteroids and of the associated olivine-dominated asteroids would require a more intense mass loss and subsequent collisional grinding. The violent collisional evolution supplied by Jupiter's formation in the case of an extensive, ≥1 au, migration of the giant planet (Turrini et al. 2012) or the one associated to the multi-planet migration phase described by the so-called "Grand Tack" scenario (Walsh et al. 2011) might then be required to explain the limited abundance of A-type asteroids in the present asteroid belt and the lack of exogenous olivine on Vesta (see Consolmagno et al. 2015 for a discussion of how such scenario compares to our post-Dawn understanding of Vesta).

**Acknowledgments**

The authors wish to thank Chris Russell and the whole Dawn team, with particular thanks to the VIR, FC, and Stereo Analysis teams. This research has been supported by the Italian Space Agency (ASI) and by the International Space Science Institute (ISSI) in Bern through the International Teams 2012 Project ''Vesta, the key to the origins of the Solar System'' ([www.issibern.ch/teams/originsolsys](www.issibern.ch/teams/originsolsys)).

*Figure 1: distribution of the olivine detections on Vesta. Concerning the detections by VIR, the yellow filled circles are the two original sites detected by Ammannito et al. (2013); the orange filled triangles are the sites detected by Ruesch et al. (2014a), i.e. the original two plus eleven additional less olivine-rich ones; the light blue filled squares are the outcrops detected by Palomba et al. (2015). Concerning the detections by FC, the green diamonds are the sites by Nathues et al. (2015). Please note that, while Ammannito et al. (2013), Ruesch et al. (2014a), and Nathues et al. (2015) provide lists of olivine-rich sites (each associated to one or more olivine outcrops), Palomba et al. (2015) provide instead a list of olivine outcrops that they group into six olivine-rich sites (based on geographical and morphological arguments) in addition to the original two. The global map of Vesta in the background has been produced by Dawn's FC and Stereo Analysis teams.*

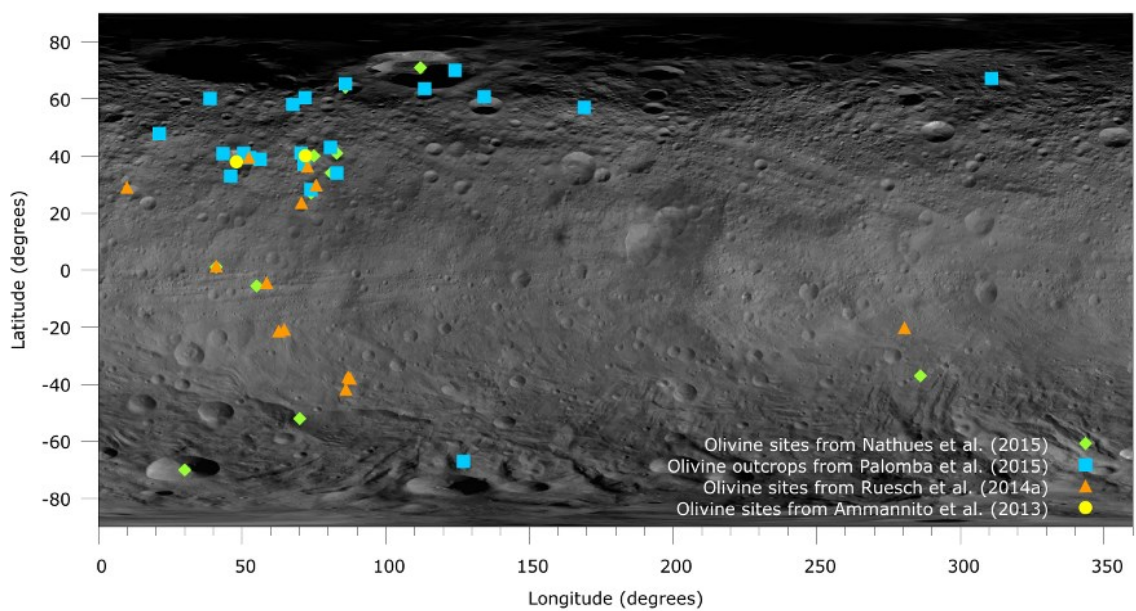

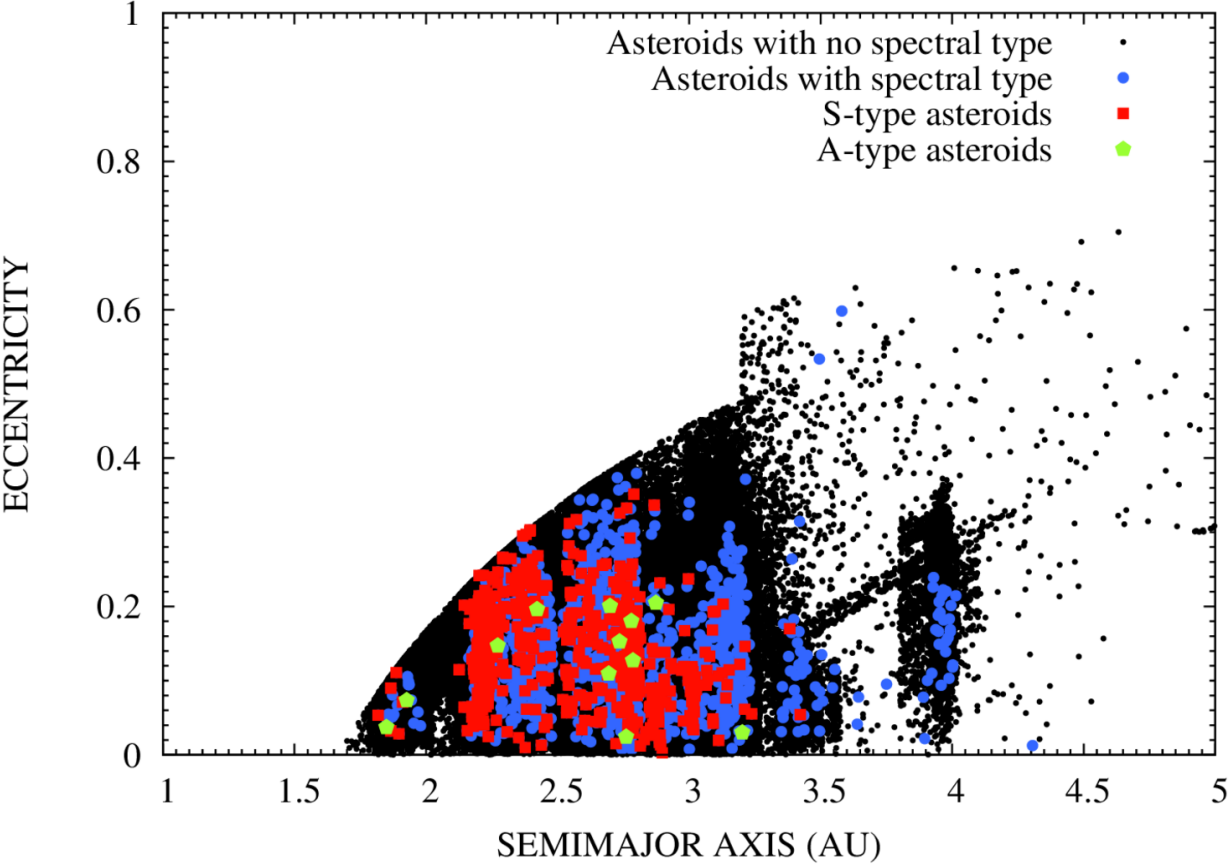

*Figure 2: orbital distribution of the asteroids in the semimajor axis - eccentricity plane. The black dots show the known asteroids without a spectral type, the blue circles the asteroids with a spectral type, the red squares the S-type asteroids, and the green pentagons the A-type asteroids. Data obtained from the JPL Small-Body Database Search Engine.*

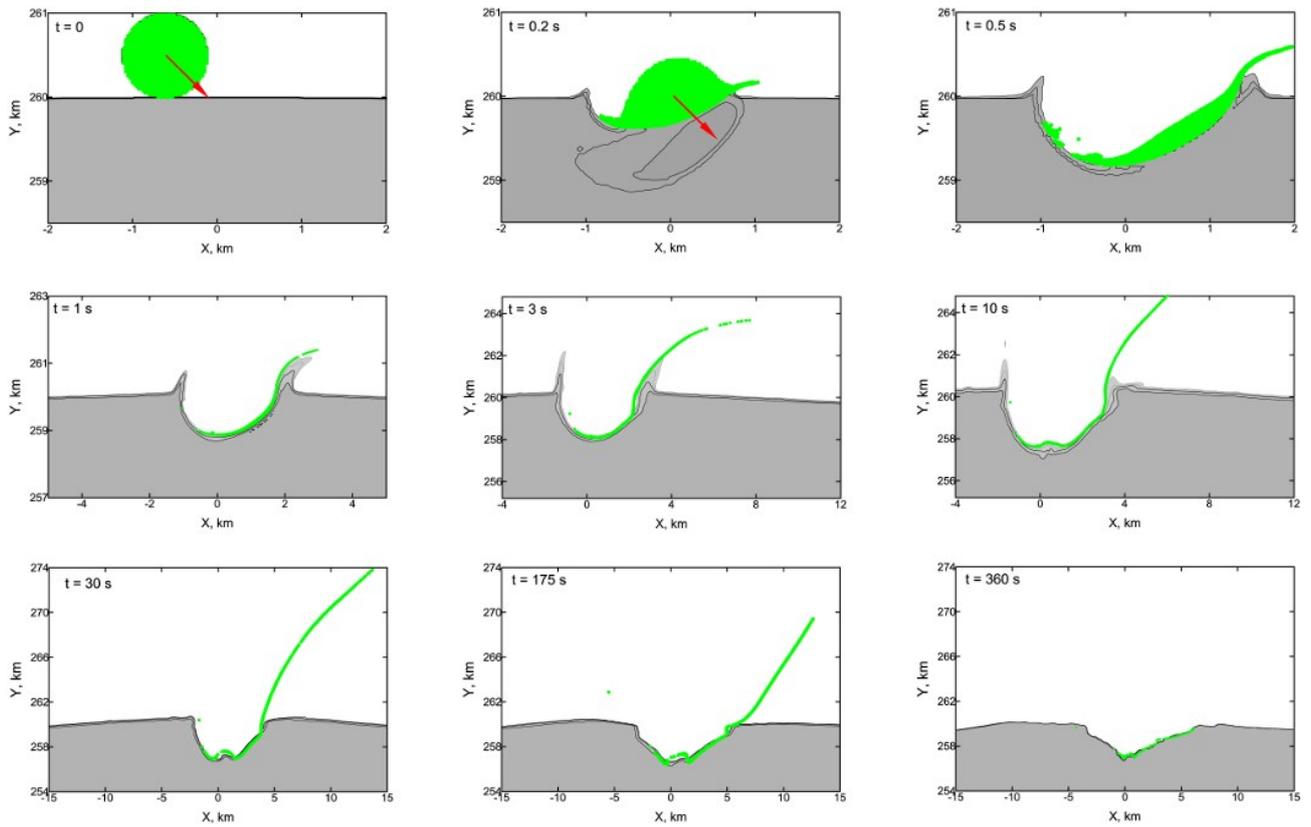

*Figure 3: Formation of the crater and motion of the material from the projectile on Vesta after the impact of a dunite planetesimal with a velocity of 4.75 km/s and a diameter of 1 km. Isolines of density are shown in the plane XY passing through the vector of impact velocity and the center of Vesta. The crustal material of Vesta with density above 0.5 g/cm$^3$ is shown by the grey color. The olivine-rich material of the impactor is shown by tracer particles in green. The red arrows in the first panels show the direction of the impact (45° respect to the local vertical).*

*Figure 4: Mass fraction of the impactor settling on Vesta's surface within a given distance from the crater center. The axis of abscissas has two scales: linear from 0 to 10 km and logarithmic from 10 to 1000 km. Panel **a** shows the results of simulations for the impacts of bodies with diameter of 1 km made from dunite and granite, a differentiated body (from dunite with an iron core of 500 m in diameter), and a body with diameter of 10 km made from dunite. The impact velocity is 4.75 km/s and the impact angle is 45°. Impactors of 1 and 10 km in diameter should produce, under these impact conditions, craters with diameters of about 10 (whose diameter is indicated by the dashed vertical line) and about 100 km, respectively. Panel **b** shows the results for the impacts of dunite bodies with diameter of 1 km with velocities 2, 4.75 and 8 km/s. After the impact at 4.75 km/s about half of the body's mass escapes, ~35-40% of its mass falls within the crater and at a short distance from its edge, and the remaining ~10-20% disperse over the surface of Vesta at large distances from the crater.*

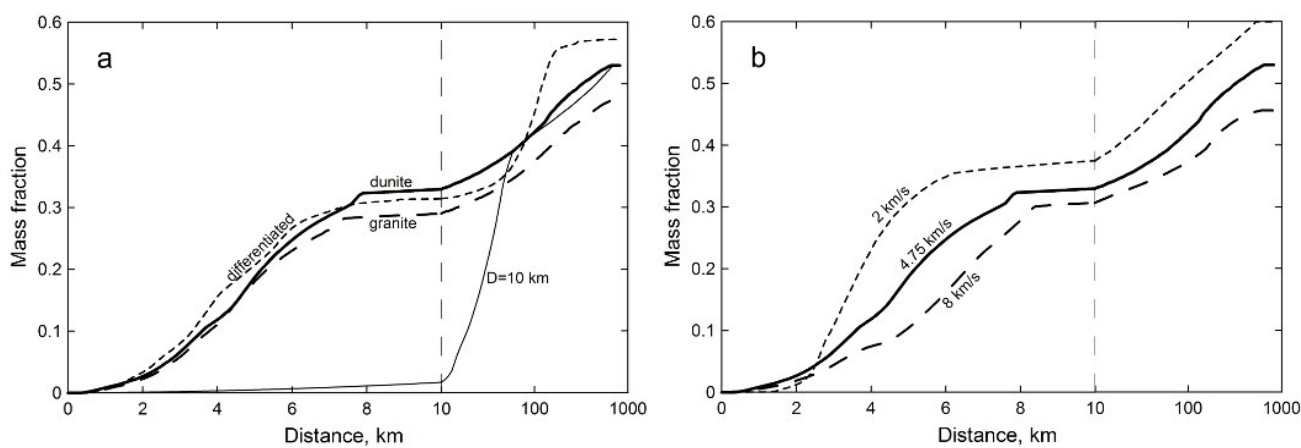

*Figure 5: Mass fraction of the impactor that settles on Vesta's surface inside a unit angle of one degree as a function of the azimuth from the crater center. Azimuth 0 corresponds to the direction of the impact velocity. The left panel **a** shows the results of simulations for the impacts of bodies with diameter of 1 km made of dunite and granite, a differentiated body (dunite with an iron core of 500 m in diameter), and a body with diameter of 10 km made of dunite. The impact velocity is 4.75 km/s and the impact angle is 45°. The right panel **b** shows the results for the impacts of bodies with diameter of 1 km made of dunite with velocities 2, 4.75 and 8 km/s.*

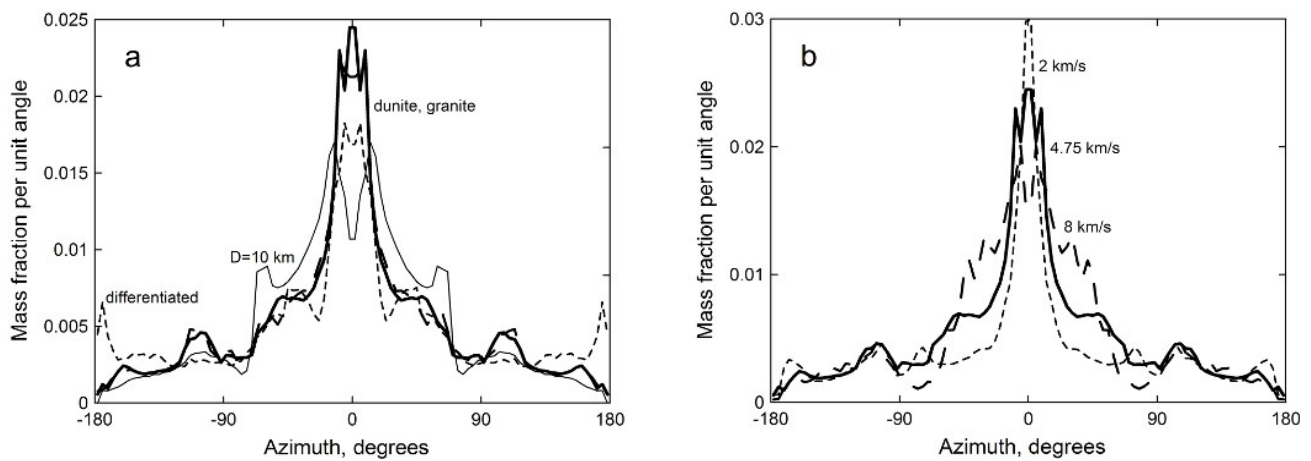

*Figure 6: Distributions over the vestan surface of the material from the projectile after the impacts of dunite planetesimals having diameters of 1 km (lower plots) and 10 km (upper plots) and a velocity of 4.75 km/s. The left panels show the hemisphere where the impacts occur, with the center of the craters in the middle of the hemisphere. The right panels show the opposite hemisphere. The impacting bodies strike Vesta along the positive direction of the X axis at 45° with respect to the local vertical. The figures show the surface density of projectile material in g/cm$^3$ in logarithmic scale. The scales are shown to the right of the figures.*

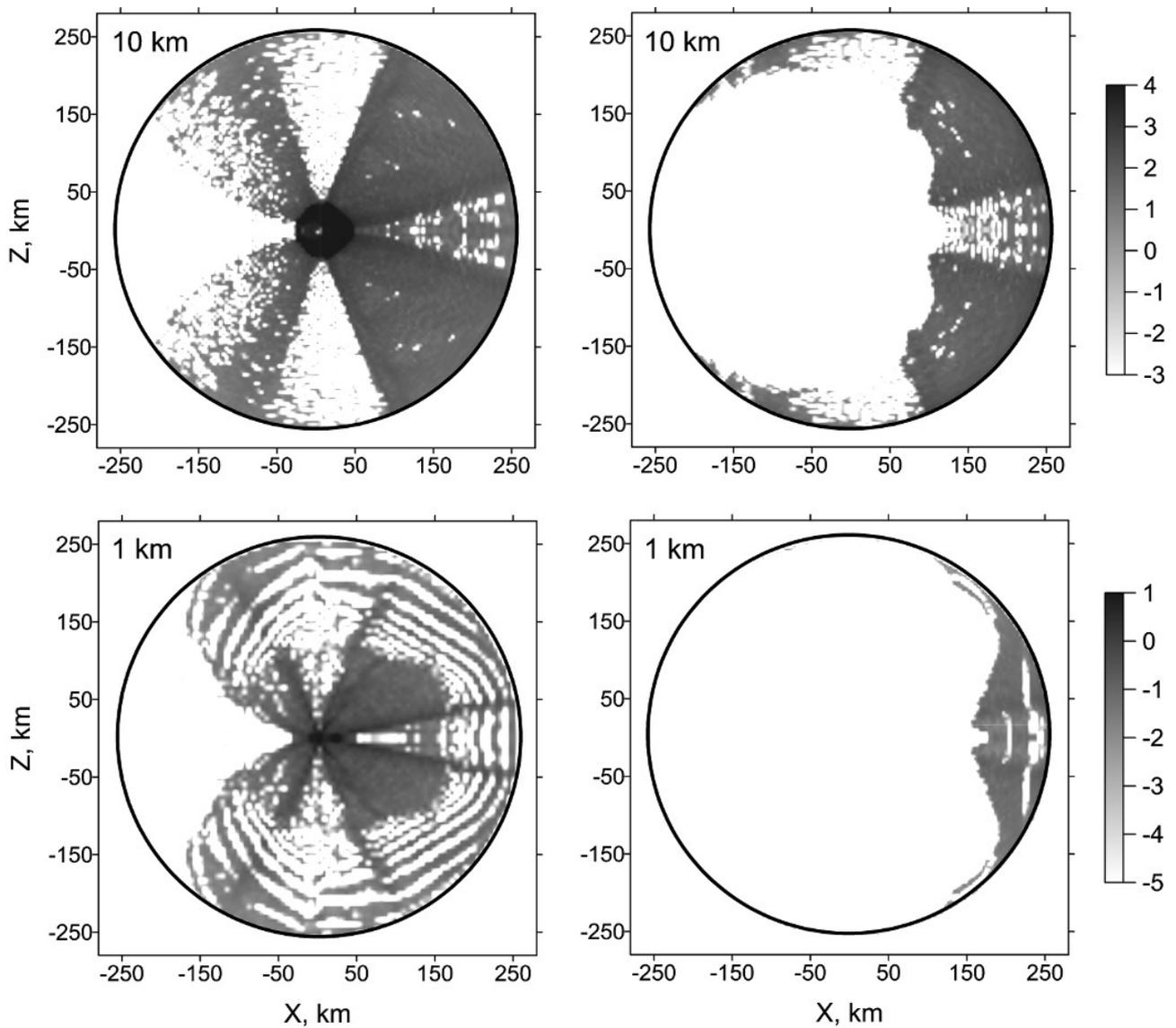